\documentclass[11pt,twoside]{article}
\usepackage{bm}
\usepackage{latexsym}
\usepackage{dcolumn}
\usepackage{amsmath,amsfonts,amssymb}
\usepackage{graphicx,epsfig}
\usepackage{psfrag}
\usepackage{amsthm}
\usepackage{color}

\usepackage{nccmath}
\usepackage{moresize}
\usepackage{enumerate}
\usepackage[hang,flushmargin]{footmisc}
\usepackage{lipsum}
\usepackage{subcaption}
\usepackage{epstopdf} 
\usepackage{hyperref}
\usepackage{rotating}
\usepackage{multirow}
\usepackage{mathtools}

\usepackage{stackengine}
\usepackage{scalerel}

\usepackage{cite}

\begin{document}

\title{\LARGE{\textbf{Reconciling Early and Late Time Tensions with Reinforcement Learning
}
}}

\author{Mohit K. Sharma\footnote{email: mr.mohit254@gmail.com} \textsuperscript{1},~ and~ M. Sami\footnote{email: samijamia@gmail.com} \textsuperscript{1,2,3} \\ 
\textsuperscript{1} {\normalsize \em Centre For Cosmology and Science Popularization, SGT University, Haryana- 122505, India} \\
\textsuperscript{2} {\normalsize \em Eurasian International Centre for Theoretical Physics, Astana, Kazakhstan} \\
\textsuperscript{3} {\normalsize \em Chinese Academy of Sciences, 52 Sanlihe Rd, Xicheng District, Beijing}
}

\date{}
\maketitle


\begin{abstract}

We study the possibility of accommodating both early and late-time tensions using a novel reinforcement learning technique. By applying this technique, we aim to optimize the evolution of the Hubble parameter from recombination to the present epoch, addressing both tensions simultaneously. To maximize the goodness of fit, our learning technique achieves a fit that surpasses even the $\Lambda$CDM model. Our results demonstrate a tendency to weaken both early and late time tensions in a completely model-independent manner.
\end{abstract}

\section{Introduction}

The recent cosmological observations related to the expansion of the universe 
and structure formation have posed significant challenges to the widely accepted 
$\Lambda$CDM model, which has long been considered the best candidate for explaining 
the universe at large scales \cite{AT-book,CST-rev,FTH-rev}. The $\Lambda$CDM model, along 
with models that mimic it by incorporating scalar degrees of freedom, is increasingly 
coming under scrutiny \cite{val-rev,sunny1}. These models often fall short of meeting the requirements set 
by recent observational data, making it a serious challenge to identify which model 
best explains various observations of the universe. Despite extensive theoretical 
efforts, a definitive answer remains elusive. However, the advent of machine learning 
(ML) offers a promising avenue to explore. In particular, it can help us to identify 
the most plausible dynamics of the universe that align more closely with observations 
than the $\Lambda$CDM model. 

Several observations, including SH0ES \cite{riess,holicow}, JWST \cite{labbe}, and the latest DESI \cite{desi}, have prompted the exploration of alternate cosmological models as they strongly disfavor 
the $\Lambda$CDM model:
\begin{itemize}
    \item \textbf{SH0ES (Supernovae $H0$ for the Equation of State)}: 
    It directly measures the Hubble constant value using SN1a (Supernovae type 1a) 
    calibrated with Cepheid variable stars. It finds $H_0=73\pm 1$\,km/s/Mpc, a 
    significant discrepancy of approximately $5\sigma$ compared to the 
    Planck 2018 CMB-derived $H_0=67.4 \pm 0.5$\,km/s/Mpc \cite{planck18}.
    \item \textbf{JWST (James Webb Space Telescope)}: 
    It measures high redshift galaxies and have found a population of surprisingly 
    massive candidates with stellar masses in the range of $10^{10}-10^{11} M_\odot$ 
    \cite{labbe,boylan}. 
    The cumulative stellar mass density of large redshift ($z=7.4\simeq9.1$) massive 
    galaxies is significantly higher than predicted by the $\Lambda$CDM model, 
    which confronts the standard model of cosmology. It has been shown that to 
    explain these observations, the star formation efficiency (SFE) needs to be at 
    least $0.57$, which is significantly higher than what previous studies have 
    reported.
    \item \textbf{DESI (Dark Energy Spectroscopic Instrument)}: 
    It takes observations of Baryonic Acoustic Oscillations (BAO) in the redshift range 
    $z\in [0.1-4.2]$ using galaxies, quasars, and Lyman-$\alpha$ as tracers. While 
    the data itself is consistent with $\Lambda$CDM, significant discrepancies arise when
    combined with other cosmological probes. This suggests a time-evolving dark energy 
    equation of state. Generalizing the $\Lambda$CDM model to $w_0w_a$CDM, DESI 
    combined with CMB and SN1a datasets gives $w_0=-0.727 \pm 0.067$ and 
    $w_a=-1.05_{-0.27}^{+0.31}$, indicating approximately $3.9\sigma$ tension with the 
    $\Lambda$CDM model \cite{desi}.
\end{itemize}

Despite the fact that observations at different redshifts are strongly challenging 
the concordance model of cosmology, there is still no clear solution of what 
phenomena are responsible for these observations or how they can be explained 
theoretically. Many efforts have been made in the literature \cite{tada,mohit-1,bousis,olga,pei,ying,dain-hTension,trodden2,periv,generic,generic2,perivola,anjan,neserris2,colgain2, piao} 
to address 
these challenges, but no common agreement on a particular explanation has been found. 
Generally, the cosmological tests on the nature of DE component is done for upto 
a redshift $z \in [0,2.5]$ only, as most data falls within this range. However, the discovery of high-redshift massive galaxies opens another door to
probe the nature of DE. The dark matter (DM) halos hosting these massive galaxies have 
time evolution and mass functions that are strongly influenced by the evolution of 
matter density perturbations $\delta_m$, which in turn depend on the evolution of the background universe \footnote{In \cite{hamsa} it was shown that keeping the background 
as $\Lambda$CDM, a Gaussian normal enhancement in the Transfer function can also 
enhance the cumulative stellar mass density.}. This dependence allows one to put constraints on the DE evolution at high redshift 
by evading the rigorous process of galaxy formation. Another crucial cosmological 
parameter in determining the limits of stellar mass content in galaxies is the matter 
density parameter. By knowing the baryonic mass fraction of the universe 
$f_b = \Omega_b / \Omega_m$, we can constrain a galaxy's stellar mass $M_\ast$ within the range defined by the product of the 
baryon fraction and the DM halo mass $M_{h}$ using the relation: 
$M_\ast \leq f_b \times M_{h}$.

In this context, we opt for a completely different strategy to search for an 
explanation for the above-mentioned observations. Specifically, we introduce a 
machine learning-based Deep Reinforcement Learning (RL) approach, which in recent 
years has shown remarkable results in various fields \cite{silver,ppo}. Our aim is to 
see whether it 
can identify underlying patterns in the observational dataset that could alleviate 
the existing tension with the $\Lambda$CDM model. The great advantage of this technique is 
that it does not require any prior form or assumption about the cosmological model. 
Being model-independent, it is a non-parametric method, which means that it does not 
need any predefined functional form to train on.

Deep Reinforcement Learning (RL) operates on a reward-based concept, where an agent 
(an explorer) interacts with an environment by taking actions at each state and 
receiving rewards based on those actions. The agent uses neural networks to modify 
its strategy in response to the rewards it receives. Once the model has sufficiently 
explored the environment to maximize rewards, it uses its optimized strategy to 
traverse states and make predictions. This approach is distinct from other 
techniques because the agent learns through dynamic interaction with the environment.

In our approach, we define the environment in terms of the Hubble parameter, from 
which we can derive all other observable quantities. By computing the maximum 
likelihood for all observations based on the form of the Hubble parameter, we treat 
this likelihood as the reward we aim to maximize. This recursive process will 
provide us with an optimized Hubble parameter profile that accounts for all 
observations equally. This method is more robust than conventional parametric 
methods, where the functional form is provided beforehand, and the task is only to 
fit parameters. The optimized result will also indicate which cosmology the model 
predicts based solely on the given data. 

The outline of the paper is as follows: We will begin by explaining the standard 
method for evaluating cumulative stellar mass density, then we will describe the 
data we used. After that, we will cover the basics of the Reinforcement Learning 
technique and how we implemented it. Finally, we will discuss the predictions that 
resulted from our training and their implications.

\section{Halo Mass Function}

The halo mass function (HMF) describes the number density of dark matter halos as a 
function of their mass. In particular, it is defined as the comoving number density 
of halos per unit mass $n(M)$. It quantifies how many halos of a given mass $M$ 
exists in a unit volume of the universe. The most general formalism that provides an 
analytical expression for the halo mass function is the Press-Schecter, which is 
based on the spherical collapse and Gaussian initial density perturbations. But 
it has some limitations i.e., it predict the over abundance around the 
characteristic mass and predict less abundance around the high mass region. 

Some alternate models such as Sheth-Tormen mass function \cite{ravi} alleviate 
these limitations by introducing ellipsoidal collapse, and gives the better fit to 
the N-body simulations. The function is given by:
\begin{equation} \label{mf}
    f(\sigma) = A \sqrt{\frac{2a}{\pi}} \left( 1 + \left( \frac{\sigma^2}{a \delta_c ^2} \right)^p \right) \frac{\delta_c}{\sigma} \exp\left(  - \frac{a \delta_c ^2}{2 \sigma^2} \right) \,,
\end{equation}
where $A= 0.3222$, $a=0.707$, $p=0.3$ are fitted parameters, and $\delta_c$ the threshold 
density contrast ($\simeq 1.686$) at which the overdense region will collapse to form a bound structure, such as halo. The variance ($\sigma$) determines the fluctuations in the density 
field smoothed over a scale corresponding to the mass $M$. It is defined by integrating 
matter power spectrum $P(k)$ over a smoothing window function \cite{mohit-pbh}, i.e. 
\begin{equation} \label{sigma}
    \sigma^2(M) = \frac{1}{2 \pi^2} \int_0^\infty P(k) W^2(k R) k^2 dk \,.
\end{equation}
Here $W(kR)$ is the spherically symmetric window function which smooth out the 
density field over a given scale $R$. In the Fourier space, it can be expressed as
\begin{equation} \label{window}
    W(k R) = \frac{3}{(k R)^3} \left[\sin(k R) - (k R) \cos(kR) \right] \,,
\end{equation}
and the matter power spectrum is given as
\begin{equation}
    P(k) = A(k) T^2(k) D^2(z) \,,
\end{equation}
where $A(k)$ is the normalization constant, $T(k)$ is the transfer function, and 
$D(z) = \delta_m(z)/\delta_m(0)$ ($\delta:=$ matter density contrast).

Thus the number density of halos in terms of $f(\sigma)$ can be written as
\begin{equation}
    \frac{dn(M,z)}{dM} = - \frac{\rho_m^{(0)}}{M} \frac{d \ln \sigma}{d M} f(\sigma) \,, 
    \qquad \mbox{such that} \qquad M = \frac{4}{3} \pi R^3 \rho_m^{(0)} \,,
\end{equation}
where $\rho_m^{(0)}$ is the present value of matter energy density. Since 
$d \ln \sigma/d M$ is negative, the minus sign ensures the number density $dn(M,z)/dM$ 
remains positive. Using above equation, the comoving number density of halos above a 
certain DM halo mass threshold ($M_{halo}$) read as \cite{boylan}
\begin{equation}
    n(>M_{halo},z) =  \int_{M_{halo}}^\infty dM \frac{dn(M,z)}{dM}
\end{equation}
and the corresponding comoving halo mass density can be written as
\begin{equation}
    \rho(>M,z) = \int_{M_{halo}}^\infty dM M \frac{dn(M,z)}{dM} \,.
\end{equation}
Depending upon the baryon fraction in the universe $f_b \equiv \Omega_b/ \Omega_m$ 
and the star formation efficiency $\epsilon \in [0,1]$, which measures how effectively gas is converted into stars, we can derive the cummulative comoving stellar mass density 
above a particular stellar mass $M_\ast$ as
\begin{equation}\label{rho_M}
    \rho(>M_\ast, z) = \epsilon f_b \int_{z_1} ^{z_2} \int _{M_\ast/\epsilon f_b} ^\infty dM M 
    \frac{dn(M,z)}{dM} \frac{dV}{V(z_1,z_2)}\,,
\end{equation}
where $M_\ast = \epsilon f_b M_{halo}$ and $V(z_1,z_2)$ is the comoving volume between 
two redshift values: $z_1$ and $z_2$. 

The observations from the JWST Cosmic Evolution Early Release
Science Survey (CEERS) program finds massive galaxies $M_\ast > 10^{10}$$M_\odot$ 
at high redshifts $z \in [7.4,9.1]$. For the observed dataset, Labbe et. al. \cite{labbe} derived 
the cumulative comoving stellar mass density and found 
it to be significantly higher than predicted by the $\Lambda$CDM model. This tension 
with $\Lambda$CDM either requires a large star formation rate or large baryon fraction 
in the collapsed structures. 

\subsection{Observational Data}\label{sec2}

For training our model, we consider the following datasets and calculate their 
respective $\chi^2$ as described below:
\begin{itemize}
    \item[(1)] \underline{\textbf{H(z)}}: We use a compilation of $48$ H(z) 
    data points obtained from different surveys from differential age 
    and galaxy clustering techniques, which ranges between redshift $z \in [0.089,2.40]$ 
    \cite{mohit-1}. 
    The corresponding $\chi^2_H$ is defined as:
    \begin{equation}
        \chi^2_H := \sum_i \left(\frac{H_\text{obs}(z_i) - H_\text{RL}(z_i)}{\sigma} \right)^2
    \end{equation}
    \item[(2)] \underline{\textbf{JWST}}: We use $4$ data points of cumulative 
    stellar mass density in two redshift bins: $z \in [7,8.5]$, and $z\in [8.5,10]$, 
    given in \cite{labbe}. However, these points are derived using the 
    Planck \texttt{TTTEEE+lowE+lensing} best-fit values for the $\Lambda$CDM model. 
    Therefore, to fit a model we must rescale the comoving volume 
    as well as luminosity distances of the given scenario to that of the $\Lambda$CDM model. 
    The $\chi^2_\text{JWST}$ is given as:
    %
%
    \begin{equation}
        \chi^2_\text{JWST} := \sum_i \left( \frac{\ln \rho_\text{th}(M_i) -  
        \ln \rho_\text{RL}(M_i)}{\sigma_\text{JWST}} \right)^2 \big|_{7<z<8.5}
        +
        \sum_i \left( \frac{\ln \rho_\text{th}(M_i) -  
        \ln \rho_\text{RL}(M_i)}{\sigma_\text{JWST}} \right)^2 \big|_{8.5<z<10} \,.
    \end{equation}
    \item[(3)] \underline{\textbf{SN1a}}: We use a collection of Supernovae type 1a 
    dataset which consists of the measurement of apparent magnitude $m_B$ in the 
    redshift range: $z \in [0.014,1.6123]$ \cite{mohit-1} \footnote{We have used the Pantheon 
    dataset for training our model because incorporating the larger Pantheon+ dataset would significantly 
    increase the training time complexity. This increase arises from the inability to parallelize the 
    RL pipeline, which would lead to longer computational time when processing the larger dataset in 
    each iteration.}. The $\chi^2_\text{SN}$ is given as 
    \footnote{Here we follow the standard procedure of marginalizing nuisance parameters, 
    such as $M_B$ and $H_0$, when calculating $\chi^2_\text{SN}$.}:
    \begin{equation}
        \chi^2_\text{SN} := \Delta m_B \cdot C^{-1}_\text{SN} \cdot \Delta m_B \,.
    \end{equation}
    where $\Delta m_B$ is the difference between observed and calculated value 
    of apparent magnitude $m_B$ at a given redshift $z$, and $C_\text{SN}$ is the 
    covariance matrix between data points.
    \item[(4)] \underline{\textbf{BAO}}: For BAO data, we use the $5$ recent DESI 
    observations between redshift $z \in [0.51, 2.33]$. Three data points at redshifts 
    $0.51$, $0.71$, and $1.32$ belong to the Luminous Red Galaxy (LRG) sample, while one data 
    point at redshift $0.93$ is part of the combined LRG and Emission Line Galaxy (ELG) sample, 
    and one data point belongs to the Ly$\alpha$ QSO sample (see Table\,(1) of \cite{desi}).
    The rest of the BAO data points are taken from \cite{bharat}. The $\chi^2_\text{BAO}$ is given as:
    \begin{equation}
        \chi^2_\text{BAO} := \Delta X \cdot C^{-1}_\text{BAO} \cdot \Delta X \,,
    \end{equation}
    where $C_\text{BAO}$ is the covariance matrix between BAO data, and $X$ represents measurement quantities such as $D_M/r_d$ and $D_H/r_d$, which are given as:
    \begin{eqnarray}
        D_M(z) &=& \frac{c}{H_0} \int_0^z dz' \, \frac{1}{E(z')}  \,,  \quad 
        \mbox{where} \quad E(z') := \frac{H(z')}{H_0} \,, \\
        D_H(z) &=& \frac{c}{H(z)}  \,.
    \end{eqnarray}
    
\end{itemize}
For these datasets, the total $\chi^2_T$ is given as:
\begin{equation}
    \chi^2_T =  \chi^2_H + \chi^2_\text{JWST} + \chi^2_\text{SN} + \chi^2_\text{BAO} \,,
\end{equation}
which is the resultant metric that we intend to minimize through our training. 
Here note that we have not considered correlations between the measurement of $H(z)$ from galaxy clustering and the result from BAO data (for more details, see 
\cite{ness-2209.12799}).

\section{RL Agent Training}

Given the large and diverse dataset spanning different redshifts, our goal is to 
obtain a model-independent expansion history of the universe without relying on any 
specific cosmological model. All the observations in the dataset ultimately depend on 
the Hubble parameter $H(z)$, which is usually chosen to explain observational phenomena. However, we aim to generate the evolution of $H(z)$ without any prior assumptions 
about its form. The training procedure is as follows:
\begin{itemize}
    \item \underline{\textbf{Setting up an Environment}}: We first set up our environment 
    for the agent to interact with and learn from. In this environment, we provide 
    the agent with a reasonably large number of possible actions, approximately $30$, 
    that it can take at any time step. The training duration for our model ranges 
    between $N = \ln(a/a_0) \in [-7.1,0]$ (from recombination to the present epoch), 
    with each time step $\Delta N = 0.05$. This small step interval allows the agent 
    to learn more fine details about the expansion history. Given the total number of steps 
    is $142$ and there are $30$ possible actions at each step, this results in a complexity 
    of around $142^{30}\sim 10^{64}$ possible states within our framework.
    
    \item \underline{\textbf{Action Space}}: At each time step $\Delta N$, the agent 
    can choose from a set of possible actions, where different actions defines 
    different fraction of change in the Hubble parameter from its previous value. 
    The transition between states are governed by the following:
    \begin{equation}
        \text{state}_{t+1}= \text{state}_t \times D(\text{action}_t)
    \end{equation}
    where $D(action_t)$ is the value associated with the action at time $t$.  

    \item \underline{\textbf{Reward Structure}}: The reward $R$ is 
    determined by a statistical test, such as Likelihood or $\chi^2$ function. 
    The goal is to maximize the reward over an episode.

    \item \underline{\textbf{Policy Update}}: The policy $\pi(a|s)$ defines 
    the probability of taking action $a$ given state $s$. The agent continuously 
    explores for optimal policy to increase the likelihood of actions that yield 
    higher rewards.
\end{itemize}

\begin{figure}[!t]
    \centering
    \includegraphics[width=0.9\linewidth,height=0.6\linewidth]{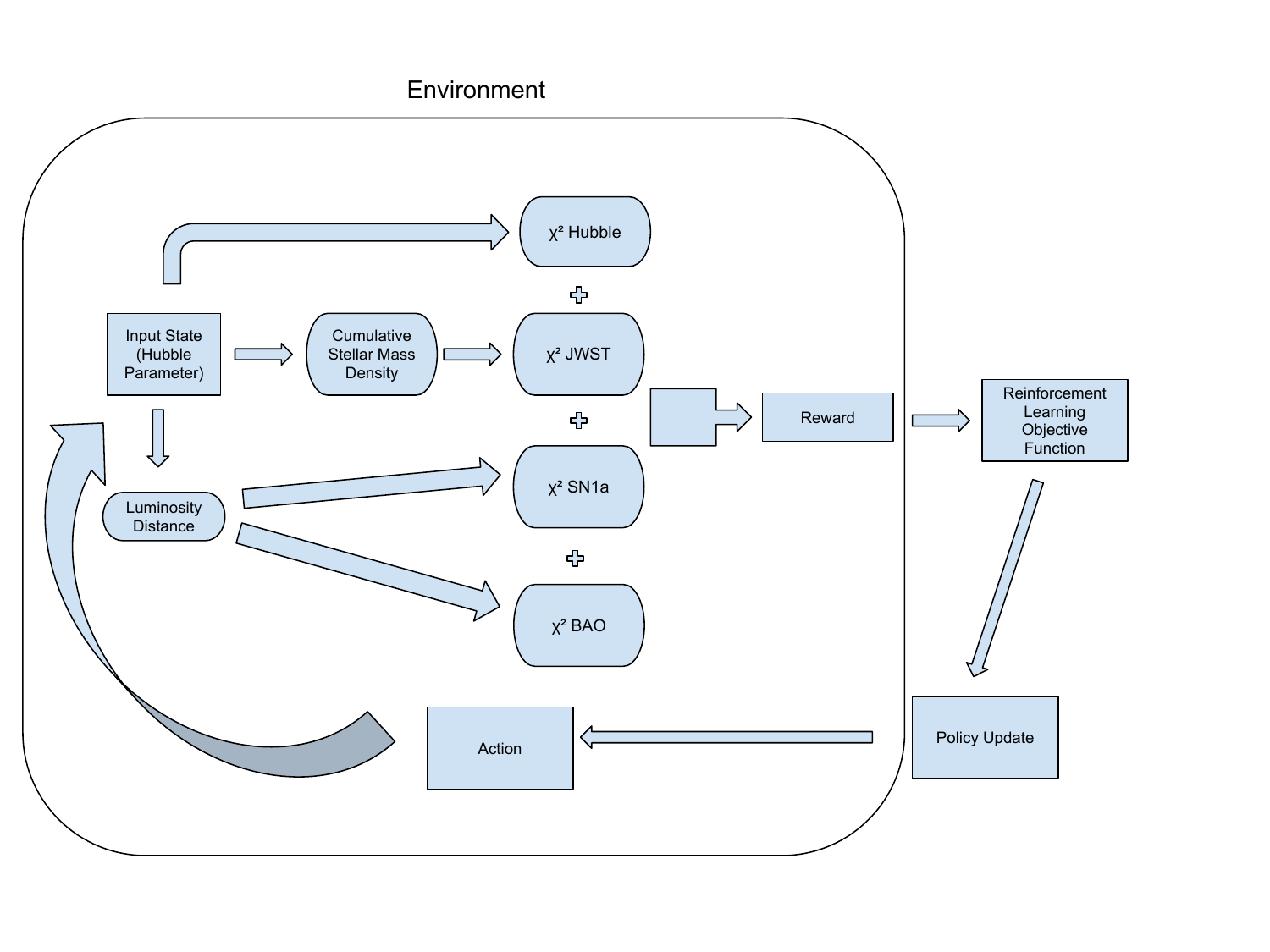}
    \caption{\sl RL framework pipeline illustrating the interaction between the environment and the 
    training objective function. The pipeline begins with the input state, which 
    corresponds to the value of the Hubble parameter as derived from the Planck best-fit 
    $\Lambda$CDM model at the recombination epoch.}
    \label{fig:flowchart}
\end{figure}

The pipeline architecture to obtain the model-agnostic best-fit scenario is 
shown in Fig. (\ref{fig:flowchart}). The architecture has two main parts: (i) the environment, 
which includes observations, rewards, and actions, and (ii) the RL algorithm, which 
tries to find the true distribution of actions given an input state.

In the environment, the state represents the possible value of the Hubble parameter 
at some time step. Initially, the algorithm explores possible states by randomly 
generating a set of states ranging from the recombination epoch to the present epoch 
\footnote{We note that the evolution of the Hubble parameter, for both training and 
predictions, begins at the recombination epoch, where its value is fixed as given by the Planck 
best-fit $\Lambda$CDM model. Subsequently, the agent selects actions from a uniform random 
distribution during the initial phases of its environment exploration.}. 
Based on each Hubble parameter's evolution, we numerically solve the second-order differential 
equation of matter density contrast $\delta_m(N)$ given by:
\begin{equation}
    \frac{d^2 \delta_m}{dN^2} + \frac{1}{2} \big[1-3 w_\text{eff}(N) \big] \frac{d \delta_m}{dN} = \frac{3}{2} \Omega_m \delta_m \,, \quad \mbox{such that} \quad 
    w_\text{eff}(N) = -1 - \frac{2}{3}\frac{H'(N)}{H(N)} \,,
\end{equation}
with the initial conditions 
$\delta_m(N) = \delta_m'(N) = 0.001$ at $N=-7$. We use the Planck $\Lambda$CDM best-fit value for 
$\Omega_m^{(0)}$ (the present matter density parameter) during both training and predictions. The reason is 
that when trying to learn the function of $H(z)$ for a particular training episode, the 
functional profile might not match the standard forms of $H(z)$ that help us estimate the parameters. 
This mismatch could lead to poor results. To avoid this, we stick with the Planck $\Lambda$CDM best-fit 
value for $\Omega_m^{(0)}$. Using $\delta_m(N)$ and the 
standard form of the Transfer function $T(k)$, we calculate the cumulative stellar 
mass density (Eq. \ref{rho_M}), which helps us get $\chi^2_\text{JWST}$. 
We also obtain luminosity distances to calculate $\chi^2_\text{SN1a}$ and 
$\chi^2_\text{BAO}$.

For each episode of our training, we calculate the sum of all $\chi^2$ values to get 
the reward. This reward information is sent to the RL objective function 
(Eq. \ref{objective-funct}). The algorithm uses a ``gradient ascent" strategy to 
maximize the reward, updating its exploration policy based on this. The policy 
determines which action should be taken at each time step. The policy is updated 
until the distribution of actions for each time step becomes stable. Once stabilized, 
the algorithm selects actions with the highest probability in a given state, 
leading to the saturation of accumulated rewards, which indicates that the model 
is now trained.

\section{Results}

\begin{figure}
    \centering
    \includegraphics{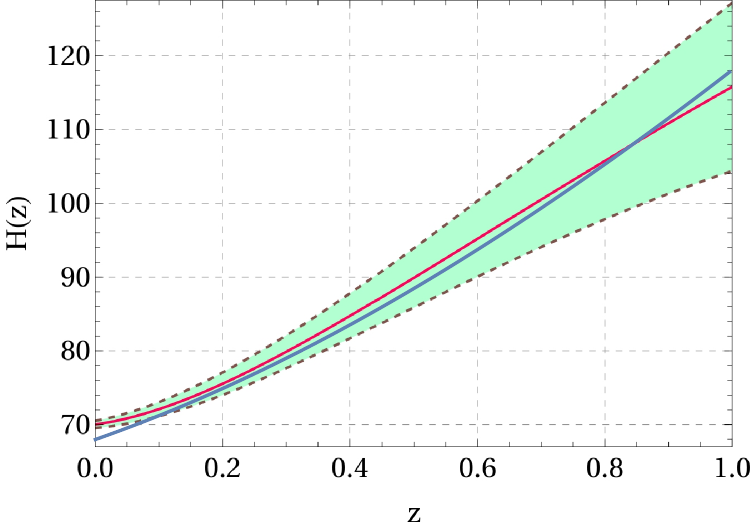}
    \caption{\sl Evolutionary profile of the Hubble parameter $H(z)$ for $z\in [0,1]$ obtained 
    using the RL framework. The solid red line represent the median 
    value, whereas dashed lines represent $1\sigma$ error bar region. The blue 
    line represents the $H(z)$ evolution for the $\Lambda$CDM model.}
    \label{fig:Hubble}
\end{figure}

Once the model is trained, it is able to select the optimized actions for 
each state. In particular, given a state-value of $H(z)$ at a particular time, 
based on the distribution of actions, it can figure out what would be the Hubble 
parameter value at next time step. The obtained evolution is shown in 
Fig.\,(\ref{fig:Hubble}), in which the solid line (best-fit) represents 
the predicted state values, and the dashed line represents the $1\sigma$ error region 
\footnote{We have calculated the error using the path integral method shown in 
\cite{ness-ga}}. 
The best-fit line corresponds to the state values of the optimized model
\footnote{Let us here mention that to create a continuous function from the discrete 
need to apply smoothing. We use the Savitzky-Golay Smoothing Filter for this purpose with 
a window length of 9 and polynomial order of 8, as described in the Appendix \ref{SG-app}.}. One 
can see that near the present epoch ($z = 0$), it predicts the Hubble parameter to be significantly 
larger than the Planck best-fit value of $67.66$ km/s/Mpc. This enhancement in 
$H_0$ can be attributed to the fact that DE could possibly be phantom in nature at late-times. 
This result is completely opposite to the DESI's combined estimates with other datasets such as 
SN1a, where it was found that DE equation of state is quite larger than 
$-1$ (as also mentioned earlier). In particular,  DESI+CBM+Pantheon+ reports $w_{DE}^{(0)} = -0.827 \pm 0.063$ 
for $w0waCDM$ parameterization, and in contrast to that we have found the DE 
equation of state parameter to be $\simeq -1.34$ when assuming $\Omega_{DE}^{(0)} = 0.7$. 
Since, in our results, it is difficult to obtain the exact functional form of $H(z)$ in terms 
of cosmological parameters, therefore, we can only quote best-fit value for the DE equation of 
state obtained using RL, assuming standard cosmological scenario. We also note that the $w_0w_a$CDM model from DESI reports 
$\Omega_m^{(0)} = 0.344^{+0.047}_{-0.026}$, which is consistent with the Planck $\Lambda$CDM 
value used in our analysis. This indicates that our choice to fix $\Omega_m^{(0)}$ to the Planck 
$\Lambda$CDM value does not contribute to any discrepancy in the present value at the present epoch 
of DE equation of state when compared with the results obtained by DESI. Since, the predicted $H(z)$ 
profile, while not accurately reconstructable using the Chevallier-Polarski-Linder (CPL) parameterization, 
in our case it is challenging to determine the precise evolution of DE equation of state. 

\begin{figure}
    \centering
    \includegraphics[width=0.5\linewidth]{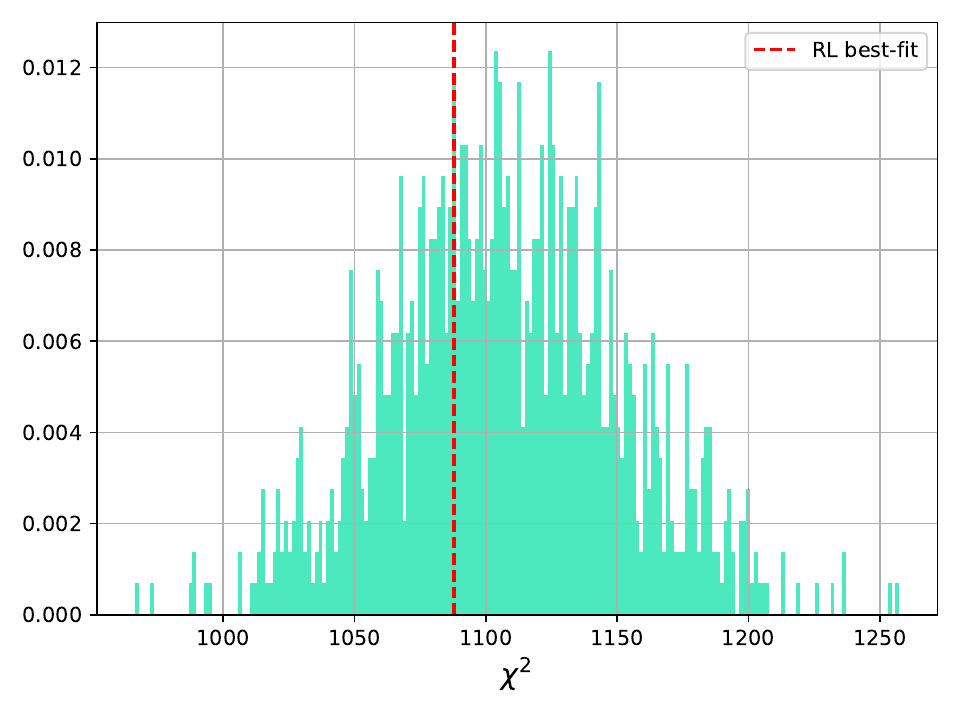}
    \caption{\sl This figure illustrates the distribution of randomly generated samples in the residual space for the $\Lambda$CDM model (without JWST). The samples are drawn from a Gaussian distribution with zero mean and diagonal covariance matrix $C$, denoted as $N(0, C)$. The red dashed line indicates the minimum $\chi^2$ value obtained via the RL method.}
    \label{fig:chi-distribution}
\end{figure}

We have observed that the $5\sigma$ tension between Planck's $\Lambda$CDM result and 
the SH0ES estimate for the Hubble constant is significantly reduced to $2.6\sigma$ through 
the RL-based reconstruction of $H(z)$ using the combined dataset. It should be noted that 
the tension still persists when applying the $\Lambda$CDM model to the combined data. 
In our approach the $H(z)$ trajectory shows closer alignment with that of the Planck's $
\Lambda$CDM at $z > 0.2$  (see Fig.,\ref{fig:Hubble}). Whereas, near the present epoch, 
it comes close towards the SH0ES findings, and thereby reduces the tension. 
Now, in order to check the goodness of fit of our result with 
respect to the $\Lambda$CDM model, we compare the minimized $\chi^2$ value for both 
cases. We find that 
\begin{equation}
   \Delta \chi^2 := \chi^2_\text{RL} - \chi^2_{\Lambda CDM} = -6.94 \,.
\end{equation}
The best-fit parameters for the $\Lambda$ CDM model are $\Omega_m^{(0)} = 0.322$ and $H_0 = 68.2$ km / s / Mpc. The negative sign of
$\Delta \chi^2$ shows the improvement of our fit compared to $\Lambda$CDM and similar models that mimic it at both early and late times. Here, note that for the combined dataset, $\Delta \chi ^2_\text{JWST} = -5.5$. It shows how well the fitted model aligns with the JWST observations. This
significant improvement in $\chi ^2_\text{JWST}$ makes it compatible with the Labbe results \cite{labbe}. To see how much the JWST data influence the
overall fit, we now remove the JWST observations and follow the same procedure using the rest of the data. We find:
\begin{equation}
  \mbox{Without JWST:} \quad  \Delta \chi^2 := \chi^2_\text{RL} - \chi^2_{\Lambda CDM} = -2.89 \,.
\end{equation}
This shows that RL-based reconstruction is still preferred over the $\Lambda$CDM model (see fig.\,(\ref{fig:chi-distribution})). However, the improvement in $\chi^2$ seen with JWST suggests that its observations support a model that deviates noticeably from $\Lambda$CDM.

\begin{figure}[t]
    \centering
    \includegraphics{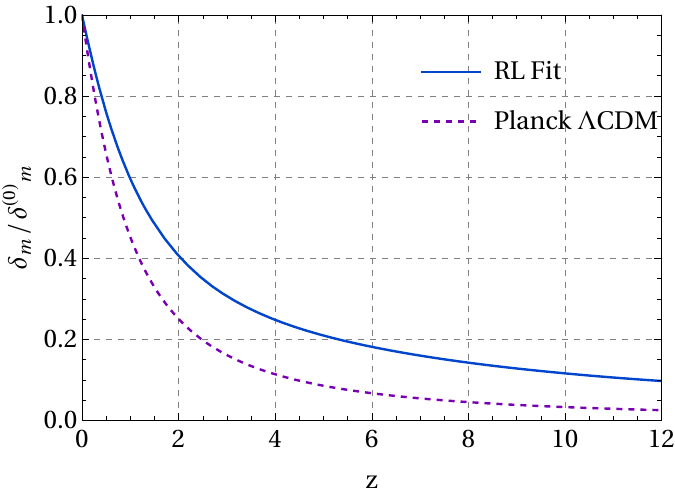}
    \caption{\sl Evolutionary profile of the normalized matter density contrast with $z \in [0, 12]$ 
    for the Planck $\Lambda$CDM best-fit (dashed) and the prediction from the RL model.}
    \label{fig:density_cont}
\end{figure}

For the obtained profile $H(z)$, we have numerically determined the evolution of the
matter density contrast $\delta_m(z)$, normalized to unity at the present epoch, as shown
in Fig.,(\ref{fig:density_cont}). In this figure, it can be observed that at higher redshifts, $D(z)$ 
or $\delta_m(z)/\delta^{(0)}_m$ obtained through RL tends to be larger at all epochs
compared to what is predicted by the Planck best fit value for the $\Lambda$CDM model. 
This occurs because phantom DE introduces more friction to the evolution of $\delta_m(z)$ 
by also decreasing the contribution of the source term. Consequently, $D(z)$ decreased comparatively slowly in the past than in the $\Lambda$CDM model. At around redshift 
$z=10$, we have found that the ratio between our best-fit $D(z)$ and 
$D(z)_{\Lambda CDM}$, i.e., $D(z)/D(z)_{\Lambda CDM}$, is $3.46$.

The observed enhancement in the matter density contrast can exponentially affect the 
cumulative stellar mass density at higher redshifts for all comoving scales. 
In particular, enhancement in $D(z)$ will enhance the matter power spectrum $P(k)$, 
which then enhance the halo mass function for a given mass $M$. This enhancement will 
then exponentially affect the cumulative stellar mass density. 
\begin{figure}[!ht]
    \centering
    \begin{subfigure}[b]{0.63\textwidth}
        \centering
        \includegraphics[width=\textwidth]{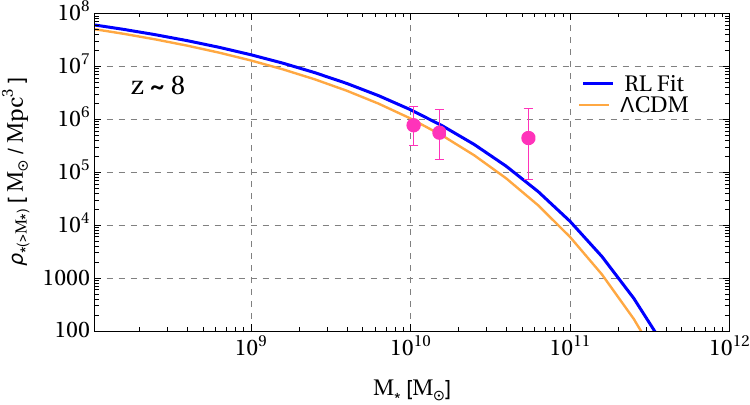}
        \caption{\sl Cumulative stellar mass density profiles at $z \simeq 8$ for both the RL 
        fitted model and the $\Lambda$CDM model, based on the combined dataset (\ref{sec2}). The three 
        data points (in purple color) refer to the JWST observations in the redshift bin $7<z<8.5$.)}
        \label{fig:rho_8}
    \end{subfigure}
    \hfill
    \begin{subfigure}[b]{0.63\textwidth}
        \centering
        \includegraphics[width=\textwidth]{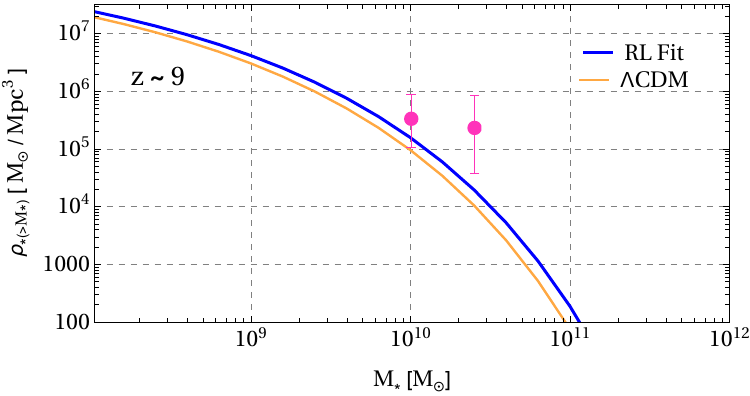}
        \caption{\sl Cumulative stellar mass density profiles at $z \simeq 8$ for both the RL 
        fitted model and the $\Lambda$CDM model, based on the combined dataset (\ref{sec2}). The two 
        data points (in purple color) refer to the JWST observations in the redshift bin $8.5<z<10$.)}
        \label{fig:rho_9}
    \end{subfigure}
    \caption{Comparative analysis of cumulative stellar mass density profiles at different redshifts. 
    \textbf{}}
    \label{fig:rho}
\end{figure}

Figs.\,(\ref{fig:rho_8}) and (\ref{fig:rho_9}) depict the cumulative stellar mass density as a function of $M_\ast$ (in solar mass $M_\odot$) for redshifts 
$8$ and $9$, respectively. These results show that the RL framework 
predicts a higher cumulative stellar mass density compared to the $\Lambda$CDM 
model. Notably, at redshift $z=9$, the $\Lambda$CDM model significantly underestimates the stellar mass density required to match observational data 
from the JWST, which suggests a need for higher densities to align with observational data.

The trained model not only suggests the reduction of the $H_0$ tension but also 
the tension with the JWST data. Since both early and late-time tensions are reduced, it indicates that the fundamental nature of DE, in overall, likely to be significantly 
differ with that of the cosmological constant.

\section{Conclusion}

In this paper, we have studied discrepancies between the early and late time observational data, such as JWST and DESI, and the underline cosmology from a completely different 
standpoint, i.e. by utilizing the model-independent reinforcement learning. As the data 
from various phenomena such as large-scale structure or BAO, which are 
inexplicable within our current understanding of theoretical models, it necessitates 
the use of a model-independent technique, specifically one that is free from any 
cosmological pre-assumptions. The main objective to implement this technique 
is to figure out if there exists any unknown feature in the data 
in the data that our conventional cosmological models are unable to take them in 
account due to pre-imposed constraints on their formulation. 

In order to estimate the statistical quantity to observe the goodness of fit, we have formulated this reward based technique in terms of $\chi^2$ which has directed the RL 
agent to choose the policy which leads to the higher cumulative reward or lower $\chi^2$ function over an episode. By changing the policy at each epoch, the model 
tries to find the optimized reward, which at the end of the training procedure 
comes out to be statistically more preferable than the $\Lambda$CDM model by a 
significant factor. 

With our constructing of the pipeline to make the agent learn the observational data 
that are at different redshifts, we have found that the trained model predicts 
underline cosmology to be  significantly distinguishable at late times due to its 
preference to the phantom behavior of DE. This is interesting in the sense that the 
model naturally finds this as the preferable DE candidate over other and it indeed 
helps in reducing atleast the late-time tension, as shown in \cite{mohit-1} using 
genetic algorithm. We have shown that this phantom nature not only milder the tensions 
that is existed between Planck $\Lambda$CDM model and SH0ES, but also tends to reduce 
the tension with the JWST observations. In particular, the consequence of this departure 
from the base $\Lambda$CDM model, gets reflected in the growth of matter perturbations, 
which shows a comparatively enhancement in the growth function of matter perturbations at 
all times upto the present epoch (\ref{fig:density_cont}). The enhanced growth 
function then acts as a key ingredient to enhance the cumulative stellar mass density 
at higher redshifts, which suggests a potential explanation to the given JWST data 
(see figs.\,(\ref{fig:rho_8} and \ref{fig:rho_9}). Also, we have observed that our 
results are in contrast to the recent DESI observations which, 
within the template of CPL ansatz, suggests a very large equation of state parameter 
for DE at the current epoch. This might be due to the 
fact of our model-independent approach in which there are no such theoretical constraints. 
Finally, it will be interesting to determine which cosmological model our RL-based fit 
closely matches, or what cosmological model can be reconstructed based on our results 
regarding the predicted evolutionary history of the universe. We are currently working on 
these lines and will try to report soon.

\section*{Acknowledgement}
We appreciate Yiying Wang and Pei Wang for their inputs during the early stages of manuscript preparation.  MS is supported by Science
and Engineering Research Board (SERB), DST, Government of India under the Grant Agreement number CRG/2022/004120 (Core Research Grant). 
MS is
also partially supported by the Ministry of Education
and Science of the Republic of Kazakhstan, Grant No.
0118RK00935, and CAS President’s International Fellowship Initiative (PIFI).

\section*{Appendix}

\subsection{Proximal Policy Optimization}\label{PPO-app}

The main working principle of PPO is to optimize the policy of selecting 
actions at each state by exploring the environment and taking feedback from it. 
The goal is to maximise the expected reward $J(\theta)$ by updating the 
policy parameter $\theta$:
\begin{equation}\label{expected_reward}
    J(\theta) = \mathrm{E} \left[ \sum_{t=0}^T \gamma^t r_t \right] \,.
\end{equation}
By using the policy gradient theorem, the rate of change of expected 
reward with the policy parameter $\theta$ can be written as:
\begin{equation}\label{policy_grad}
    \nabla_\theta J(\theta) = \mathrm{E}_{s \sim \rho^\pi, a \sim \pi_\theta} 
   \left[ \nabla_\theta \log \pi_\theta (a|s) \hat{A}(s,a) \right] \,,
\end{equation}
where $\hat{A}(s,a)$ is the advantage function.
It is based on the gradient of the accumulated reward $J(\theta)$ with 
respect to the policy parameter $\theta$ as:
\begin{equation}
   \nabla_\theta J(\theta) = \mathrm{E}_{s \sim \rho^\pi, a \sim \pi_\theta} 
   \left[ \nabla_\theta \log \pi_\theta (a|s) \hat{A}(s,a) \right] \,.
\end{equation}

PPO uses the concept of clipped objective to ensure that the updates to the 
policy are not too large. The clipped surrogate objective is defined as:
\begin{equation}
    L^{\text{CLIP}}(\theta) = \mathrm{E}_t \left[ min\{r_t(\theta) \hat{A}_t,
    clip\{r_t(\theta), 1-\epsilon, 1+\epsilon \} \hat{A}_t \}
    \right] \,,
\end{equation}
where $r_t(\theta) = \frac{\pi_\theta(a_t|s_t)}{\pi_{\theta_{old}}(a_t|s_t)}$ 
is the probability ratio, and $\epsilon \simeq 0.2$ is an hyperparameter that 
controls the clipping range.

The policy parameter $\theta$ are updated to maximise the clipped surrogate objective 
as:
\begin{equation}
    \theta \leftarrow \theta + \alpha \nabla_\theta L^{\text{CLIP}}(\theta) \,, 
    \quad \mbox{where} \quad \alpha:=  \text{Learning rate} \,.
\end{equation} 
It also uses the value function $V_\theta (s_t)$ to estimate the expected return. 
The value function loss is defined as:
\begin{equation}
    L^\text{VF}(\theta) = \mathrm{E}_t \left[ (V_\theta (s_t) - R_t)^2 \right] \,.
\end{equation}
To encourage exploration, the Entropy is also added to the objective function:
\begin{equation}
    L^\text{S}(\theta) = \mathrm{E}_t \left[ \mathcal{H}(\pi_\theta(\cdot|s_t)) \right] \,.
\end{equation}
Finally the total objective function is given as:
\begin{equation} \label{objective-funct}
    L(\theta) = \mathrm{E}_t \left[L^{\text{CLIP}}(\theta) - c_1 L^\text{VF}(\theta) 
    + c_2 L^\text{S}(\theta) \right] \,,
\end{equation}
where $c_1$ and $c_2$ are some coefficients.

\subsection{Savitzky-Golay Smoothing Filter}\label{SG-app}

The Savitzky-Golay filter is a filtering technique that is used to smooth data 
while preserving the shape and important details in the data. For a given set 
of data points $y_i$, where $i \in [0,N-1]$, the smoothed value $\hat{y}_i$ 
s obtained by fitting a polynomial of order $p$ over a window of length $2m+1$ 
centered around each point. The formula for the smoothed value is:
\begin{equation}
    \hat{y}_i = \sum_{j=-m}^m c_j y_{i+j} \,,
\end{equation}
where $c_j$ are the filter coefficients, and $y_{i+j}$ are the original data points 
within the window centered at $y_i$. The polynomial $P(x)$ of degree $p$ can be 
written as:
\begin{equation}
    P(x) = a_0 + a_1 x + a_2 x^2 \cdot + a_p x^p \,.
\end{equation}
For every data point $y_i$, it choose a window centered on $y_i$ and fit the 
above polynomial to the points: $\{y_{i-m}, y_{i-m+1} \dots y_i \dots y_{i+m-1}, y_{i+m}\}$. 

To construct the Vandermonde matrix $A$, it use the relative positions within the window. 
If the window has length $2m+1$ and is centered around $y_i$, the relative positions 
are $k=-m,-m+1,\dots,0,\dots,m-1,m$. The Vandermonde matrix $A$ is:
$
\begin{bmatrix}
(-m)^0 & (-m)^1 & (-m)^2 & \cdots & (-m)^p \\
(-m+1)^0 & (-m+1)^1 & (-m+1)^2 & \cdots & (-m+1)^p \\
\vdots & \vdots & \vdots & \ddots & \vdots \\
0^0 & 0^1 & 0^2 & \cdots & 0^p \\
\vdots & \vdots & \vdots & \ddots & \vdots \\
(m-1)^0 & (m-1)^1 & (m-1)^2 & \cdots & (m-1)^p \\
m^0 & m^1 & m^2 & \cdots & m^p \,. \\
\end{bmatrix}
$

In the abaove matrix, each row corresponds to a data point within the window, and each column corresponds 
to a power of the relative position from the chosen point. Now, the filter coefficients $c_j$ can be obtained using the least squares: $Aa=y$, where $y$ is the vector of data points 
in the window and $a$ is the vector of polynomial coefficients. The coefficients $c_j$ 
are derived from the first row of the pseudoinverse of the Vandermonde matrix $A$.

The smoothed value $\hat{y}_i$ is then given by:
\begin{equation}
    \hat{y}_i = \sum_{j=-m}^{m} c_j y_{i+j} \,.
\end{equation}


\begin{thebibliography}{99}
\small
\bibitem{AT-book}
L. Amendola and S. Tsujikawa, {\em Dark Energy: Theory and Observations}, 
Cambridge University Press, United Kingdom (2010).
%
\bibitem{CST-rev}
E.J. Copeland, M. Sami and S. Tsujikawa, {\em Dynamics of dark energy}, 
Int. J. Mod. Phys. {\bf D 15} (2006) 1753, arXiv:\,hep-th/0603057.
%
\bibitem{FTH-rev}
J.A. Frieman, M.S. Turner and D. Huterer, {\em Dark energy and the 
accelerating universe}, Ann. Rev. Astron. Astrophys. {\bf 46} (2008) 385, 
arXiv:\,0803.0982[astro-ph].
%
\bibitem{val-rev}
E. Di Valentino et al., {\em In the realm of the Hubble tension—a review of solutions}, Class. Quant. Grav. 
{\bf 38} (2021) 15, 153001, 
arXiv:\,2103.01183 [astro-ph.CO].
%
\bibitem{sunny1}
S. Vagnozzi, {\em Seven hints that early-time new physics alone is 
not sufficient to solve the Hubble tension}, Universe {\bf 9} (2023) 393, 
arXiv:\,2308.16628 [astro-ph.CO].
%
\bibitem{riess}
A. G. Riess et. al., {\em A 2.4\% Determination of the Local Value of the Hubble Constant}, Astrophys. J., {\bf 826}(1) (2016) 56, arXiv:\,1604.01424 [astro-ph.CO]. 
%
\bibitem{holicow}
K. C. Wong et. al., {\em H0LiCOW \textendash{} XIII. A 2.4 per cent measurement of H0 from lensed quasars: 5.3\ensuremath{\sigma} tension between early- and late-Universe probes}, Mon. Not. Roy. Astron. Soc., {\bf 498}(1) (2020) 1420-1439, arXiv:\,1907.04869 [astro-ph.CO].
%
\bibitem{planck18}
N. Aghanim et. al., {\em Planck 2018 results. VI. Cosmological parameters}, Astron. Astrophys. {\bf 641} (2020) A6, arXiv:\,1807.06209 [astro-ph.CO].
%
\bibitem{boylan}
M. Boylan-Kolchin, {\em Stress testing \ensuremath{\Lambda}CDM with high-redshift galaxy candidates}, Nature Astron. {\bf 7}(6) (2023) 731--735,arXiv:\,2208.01611 [astro-ph.CO]. 
%
\bibitem{labbe}
I. Labbe et. al., {\em A population of red candidate massive galaxies 
\textasciitilde{}600 Myr after the Big Bang}, Nature {\bf 616}(7956) (2023) 266--269, 
arXiv:\,2207.12446 [astro-ph.GA].
%
\bibitem{desi}
A.G. Adame et. al., {\em DESI 2024 VI: Cosmological Constraints from the Measurements 
of Baryon Acoustic Oscillations}, arXiv:\,2404.03002 [astro-ph.CO].
%
\bibitem{tada}
Y. Tada, T. Terada, {\em Quintessential interpretation of the evolving dark energy in light of DESI observations}, Phys. Rev. {\bf D} {\bf 109}(12) (2024) L121305, 
arXiv:\,2404.05722 [astro-ph.CO].
%
\bibitem{mohit-1}
M. R. Gangopadhyay, M. Sami, M. K. Sharma, {\em Phantom dark energy as a natural selection of evolutionary processes a\textasciicircum{} la~genetic algorithm and cosmological tensions}, Phys. Rev. {\bf D} {\bf 108}(10) (2023) 103526, arXiv:\,2303.07301 [astro-ph.CO].
%
\bibitem{bousis}
D. Bousis, L. Perivolaropoulos, {\em Hubble tension tomography: BAO vs SnIa distance tension}, arXiv:\,2405.07039 [astro-ph.CO].
%
\bibitem{olga}
M. Forconi, W. Giarè, O.Mena, Ruchika, E. Di Valentino, {\em A double take on early and interacting dark energy from JWST}, JCAP {\bf 05} (2024) 097, arXiv:\,2312.11074 
[astro-ph.CO].
%
\bibitem{pei}
P. Wang et. al., {\em Exploring the Dark Energy Equation of State with JWST}, 
arXiv:\,2307.11374 [astro-ph.CO].
%
\bibitem{anjan}
S. A. Adil et. al., {\em Dark energy in light of the early JWST observations: case for a negative cosmological constant?}, JCAP {\bf 10} (2023) 072, arXiv:\,2307.12763 
[astro-ph.CO].
%
\bibitem{ying}
Yi-Ying Wang et. al., {\em Modeling the JWST High-redshift Galaxies with a General Formation Scenario and the Consistency with the \ensuremath{\Lambda}CDM Model}, 
Astrophys. J. Lett. {\bf 954}(2) (2023) L48, arXiv:\,2307.12487 [astro-ph.GA]. 
%
\bibitem{dain-hTension}
M. G. Dainotti, G. Bargiacchi, M. Bogdan, 
S. Capozziello, S. Nagataki, {\em Reduced uncertainties up to $43\%$ on the Hubble constant and the matter density with the SNe Ia with a new statistical analysis}, arXiv:\,2303.06974 [astro-ph.CO].
%
\bibitem{trodden2}
J. Sakstein, and M. Trodden, {\em Early Dark Energy from Massive Neutrinos as a Natural Resolution of the Hubble Tension}, Phys. Rev. Lett. {\bf 124}(16) (2020) 161301, arXiv:\,1911.11760 [astro-ph.CO].
%
\bibitem{periv}
G. Alestas, L. Kazantzidis, and L. Perivolaropoulos, 
{\em $H_0$ tension, phantom dark energy, and cosmological parameter degeneracies}, Phys. Rev. {\bf D} {\bf 101}(12) (2020) 123516, 
arXiv:\,2004.08363 [astro-ph.CO].
%
%
\bibitem{neserris2}
R. Arjona, and S. Nesseris, {\em What can Machine Learning tell us about the background expansion of the Universe?}, Phys. Rev. {\bf D} {\bf 101}(12) 
(2020) 123525, arXiv:\,1910.01529 [astro-ph.CO].
%

\bibitem{colgain2}
E. Ó. Colgáin et. al., {\em Do high redshift QSOs and GRBs corroborate JWST?}, 
arXiv:\,2405.19953 [astro-ph.CO].
%

\bibitem{piao}
H. Wang, Y. Piao, {\em Dark energy in light of recent DESI BAO and Hubble tension}, 
arXiv:\,2404.18579 [astro-ph.CO].
%

\bibitem{generic}
S. A. Adil, M. R. Gangopadhyay, M. Sami, and M. K. Sharma, {\em Late-time acceleration due to a generic modification of gravity and the Hubble tension}, Phys. Rev. {\bf D} {\bf 104}(10) (2021) 103534, arXiv:\,2106.03093 [astro-ph.CO].
%
\bibitem{generic2}
M. R. Gangopadhyay, S. K. J. Pacif, M. Sami, and M. K. Sharma, {\em Generic Modification of Gravity, Late Time Acceleration and Hubble Tension}, Universe {\bf 9}(2) (2023) 83, arXiv:\,2211.12041 [gr-qc].
%
\bibitem{perivola}
G. Alestas, and L. Perivolaropoulos, {\em Late-time approaches to the Hubble tension deforming H(z), worsen the growth tension}, Mon. Not. Roy. Astron. Soc. {\bf 504}(3) (2021) 3956-3962, arXiv:\,2103.04045 [astro-ph.CO].
%
\bibitem{silver}
H. Van Hasselt, A. Guez, D. Silver, {\em Deep reinforcement learning with double q-learning}, Proceedings of the AAAI conference on artificial intelligence, {\bf 30}(1) 
(2016), arXiv:\,1509.06461 [cs.LG].
%
\bibitem{ppo}
J. Schulman, et al. {\em Proximal policy optimization algorithms}, arXiv:1707.06347.
%
\bibitem{ravi}
R. K. Sheth, H.J. Mo, G. Tormen, {\em Ellipsoidal collapse and an improved model for the number and spatial distribution of dark matter haloes}, Mon. Not. Roy. Astron. Soc. 
{\bf 323} (2001) 1, arXiv:\,astro-ph/9907024.
%
\bibitem{mohit-pbh}
M. K. Sharma, M. Sami, D. F. Mota, {\em Generic Predictions for Primordial Perturbations and their implications}, arXiv:\,2401.11142 [astro-ph.CO].
%
\bibitem{bharat}
S. Cao, J. Ryan, B. Ratra, {\em Using Pantheon and DES supernova, baryon acoustic oscillation, and Hubble parameter data to constrain the Hubble constant, dark energy dynamics, and spatial curvature}, Mon. Not. Roy. Astron. Soc. {\bf 504}(1) (2021) 
300-310, arXiv:\,2101.08817 [astro-ph.CO].
%
\bibitem{ness-ga}
S. Nesseris, J. García-Bellido, {\em Comparative analysis of model-independent methods 
for exploring the nature of dark energy}, Phys. Rev. {\bf D} {\bf 88}(6) (2013) 
063521, arXiv:\,1306.4885 [astro-ph.CO].
%
\bibitem{ness-2209.12799}
G. Alestas, L. Kazantzidis, S. Nesseris, {\em Machine learning constraints on deviations from general relativity from the large scale structure of the Universe}, 
Phys. Rev. {\bf D} {\bf 106}(10) (2022) 103519, arXiv:\,2209.12799 [astro-ph.CO].
%
\bibitem{hamsa}
H. Padmanabhan, A. Loeb, {\em Alleviating the Need for Exponential Evolution of JWST Galaxies in $10^{10}$M$_{\odot}$ Haloes at z \ensuremath{>} 10 by a Modified \ensuremath{\Lambda}CDM Power Spectrum}, Astrophys. J. Lett. {\bf 953}(1) (2023) L4, arXiv:\,2306.04684 [astro-ph.CO].





\end{thebibliography}
\end{document}